# Quantum Sensing MRI for Noninvasive Detection of Neuronal Electrical Activity in Human Brains

***Short title:*** Quantum sensing MRI for humans

Yongxian Qian,[1,2,3,*] Ying-Chia Lin,[1,2] Sara Hejazi,[1,2] Kamri Clarke,[1,2] Kennedy Watson,[1,2] Xingye Chen,[1,4] Nahbila-Malikha Kumbella,[1] Justin Quimbo,[1] Abena Dinizulu,[5] Simon Henin,[6] Yulin Ge,[1,2] Arjun Masurkar,[5] Anli Liu,[6] Yvonne W. Lui,[1,2] Fernando E. Boada[1,2,†]

[1] Bernard and Irene Schwartz Center for Biomedical Imaging, Department of Radiology, New York University Grossman School of Medicine, New York, New York 10016, USA.

[2] Center for Advanced Imaging Innovation and Research (CAI2R), Department of Radiology, New York University Grossman School of Medicine, New York, New York 10016, USA.

[3] Neuroscience Institute, Department of Neuroscience and Physiology, New York University Grossman School of Medicine, New York, New York 10016, USA.

[4] Vilcek Institute of Graduate Biomedical Sciences, New York University Grossman School of Medicine, New York, New York 10016, USA.

[5] Alzheimer's Disease Research Center, Department of Neurology, NYU Langone Health, New York, New York 10016, USA.

[6] Comprehensive Epilepsy Center, Department of Neurology, NYU Langone Health, New York, New York 10016, USA.

† Present address: Department of Radiology, Stanford University, Stanford, California 94305, USA

* **Corresponding author**. Email: Yongxian.Qian@nyulangone.org

## Abstract

Neuronal electrical activity underlies human cognition, yet its direct, noninvasive measurement in the living human brain remains a fundamental challenge. Existing neuroimaging techniques, including EEG, MEG, and fMRI, are limited by trade-offs in sensitivity and spatial or temporal resolution. Here we propose quantum sensing MRI (qsMRI), a noninvasive approach that enables direct detection of neuronal firing-induced magnetic fields using a clinical MRI system. qsMRI exploits endogenous proton ($^1$H) nuclear spins in water molecules as intrinsic quantum sensors and decodes time-resolved phase information from free induction decay (FID) signals to infer neuronal magnetic fields. We validate qsMRI through simulations, phantom experiments, and human studies at rest and during motor tasks, and provide open experimental procedures to facilitate independent rigorous validation. We further present a case study demonstrating potential applications to neurological disorders. qsMRI represents, to our knowledge, the first-in-human application of quantum sensing on a clinical MRI platform and may lay the foundation for a non-BOLD functional imaging modality capable of probing neuronal firing dynamics in both cortical and deep brain regions.

**Acknowledgments**

The authors would like to thank colleagues at the Center for Biomedical Imaging at NYU Radiology and peers in the International Society of Magnetic Resonance in Medicine (ISMRM) for their thoughtful discussions and critical comments on quantum sensing and neuronal firing detection.

**Funding:**

National Institutes of Health (NIH) grants RF1/R01 AG067502 (YQ, YL, SH, KC, KW, XC, LA, NMK, JQ, AD, SH, YG, AM, AL, YWL, FEB).

Department of Radiology of the New York University Grossman School of Medicine General Research Fund (YQ).

**NIH-sponsor required statement:**

Research reported in this publication was supported in part by the National Institute On Aging (NIA) of the National Institutes of Health (NIH) under Award Number RF1/R01 AG067502. The content is solely the responsibility of the authors and does not necessarily represent the official views of the National Institutes of Health.

This work was performed under the rubric of the Center for Advanced Imaging Innovation and Research ($CAI^2R$), a National Institute of Biomedical Imaging and Bioengineering (NIBIB) Biomedical Technology Resource Center grant NIH P41 EB017183.

**Author contributions:**

Conceptualization: YQ

Methodology: YQ, YL, YWL, FEB

Investigation: YQ, YL, SH, KC, KW, XC, LA, NMK, JQ, AD, SH, YG, AM, AL, YWL, FEB

Visualization: YQ, YL, YWL

Supervision: YQ, YWL, FEB

Writing – original draft: YQ

Writing – review & editing: YQ, YL, SH, KC, KW, XC, LA, NMK, JQ, AD, SH, YG, AM, AL, YWL, FEB

**Competing interests:**

YQ, YL, XC, AL, YWL, and FEB are the inventors of the U.S. Patent Application, No.: 63/467,482. Title: *Systems and methods for non-invasive detection of neuronal firings in humans via quantum-sensing magnetic resonance imaging*. Filed on May 18, 2023. IP Owner: New York University.

All other authors declare they have no competing interests.

**Data and materials availability:**

Upon written request to the corresponding author, Yongxian Qian, PhD, all data (MRI images and FID signals) and codes used in this work (main text and supplementary materials) are available to any researcher solely for scientific research purposes.

## Main

Neuronal electrical activity (called firings), manifested as action potentials propagating along axons and across synapses, underlies communication within neural circuits that support cognition, including perception, attention, memory, and language. Disruption of these firings by aging or neurological disease impairs circuit function and leads to cognitive deficits such as memory loss, language dysfunction, and impaired decision-making – hallmarks of normal aging, mild cognitive impairment (MCI), and Alzheimer's disease (AD) and other types of dementia. Direct measurement of neuronal firings is therefore essential for elucidating the neural basis of cognition and its decline, and for improving diagnosis and treatment of neurological disorders. However, achieving noninvasive, in vivo detection of neuronal firings in humans remains a major unresolved challenge in neuroscience and clinical neurology.

Neuronal firings occur deep within the skull, and their direct measurement faces three fundamental challenges. First, the associated magnetic fields are extremely weak, typically below 0.2 pT when measured at the scalp. Second, neuronal firings are transient, with action potentials (APs) lasting ~2 ms and postsynaptic potentials (PPs) ~10–100 ms. Third, any viable approach must be noninvasive, safe, and suitable for repeated use in humans. Together, these constrains severely limit the ability of current noninvasive technologies – including electroencephalography (EEG), magnetoencephalography (MEG), and functional magnetic resonance imaging (fMRI) – to directly detect neuronal firings.

Scalp EEG and MEG have substantially advanced the study of neurological disorders, including epilepsy, brain injury, and cognitive impairment; however, they measure neuronal electric and magnetic fields from a relatively large distance (~20 mm) from their sources and therefore have limited source localization accuracy. Consequently, their sensitivity is largely restricted to slow, highly synchronized postsynaptic activity arising from relatively large cortical regions.

Functional MRI (fMRI) maps large-scale brain networks at rest and probes task-related neural circuits using external stimuli. Although fMRI has transformed the study of human cognition, it suffers from two fundamental limitations.[1] First, it measures neuronal activity indirectly via blood oxygen level-dependent (BOLD) signals, whose relationship to underlying neuronal firing is complex and not fully understood.[2] Second, its temporal resolution (~ 40–100 ms) is insufficient to capture the rapid dynamics of action potentials and postsynaptic potentials that constitute neuronal firings.[3]

To overcome these limitations, researchers, including our group, have combined fMRI, which offers high spatial resolution (~4 mm), with scalp EEG, which provides high temporal resolution (~2 ms). However, this hybrid approach faces intrinsic challenges, including substantial electromagnetic interference between the two modalities and reduced subject comfort.[4-6] To mitigate these difficulties, alternative strategies have sought to detect oscillatory brain activity rather than action potential spikes, for example by exploring double resonance effects using spin-lock techniques,[7,8] but these methods have not yet been fully validated in humans. More recently, approaches aiming to achieve higher temporal resolutions (milliseconds or even sub-milliseconds [9]) with fMRI have been proposed, but they rely on the assumption that neuronal activity is perfectly repeatable over time – an assumption that is unlikely to hold in practical experimental settings

Here, we address these challenges by proposing quantum sensing MRI (qsMRI), a noninvasive method that leverages the established hardware of a clinical 3T MRI system equipped with a multi-channel array coil to perform quantum sensing [10] of neuronal firings in humans at room temperature. In addition to maintaining the safety and noninvasiveness of conventional MRI, qsMRI incorporates three conceptual advances. First, to detect extremely weak neuronal magnetic field (<0.2 pT at the scalp), qsMRI repurposes endogenous proton ($^1H$) nuclear spins in water molecules as intrinsic quantum sensors. These nuclear spins reside within axons, where neuronal currents generate strong dipolar magnetic fields, in contrast to the much weaker quadrupolar fields outside axons. The resulting intra-axonal dipolar magnetic field can reach ~180 μT (see below), more than nine orders of magnitude stronger than the fields detected by MEG at the scalp. Second, to capture the rapid dynamics of neuronal firing, qsMRI exploits standard MRI excitation and readout hardware to *excite* the quantum sensors (proton nuclear spins) and acquire free induction decay (FID) signals from the excited quantum sensors with sub-millisecond temporal resolution (0.01–0.20 ms), depending on the required dynamic range. Third, qsMRI localizes firing sources using either coil-element-specific sensitivity profiles or selective radiofrequency (RF) excitation of individual voxels, enabling definitive spatial attribution of neuronal activity. Together, these strategies may define qsMRI as a new class of non-BOLD functional MRI that enables direct, time-resolved detection of neuronal firings.

## Quantum sensing MRI

As conceptually illustrated in Figure 1, qsMRI operates on a clinical 3T MRI system equipped with a multi-channel array coil. The method comprises two core steps: localization of neuronal firing sources and recording of neuronal firings. Firing sources are localized using either a passive or an active strategy (Figure 1a). In the passive strategy, an imaging pulse sequence, such as scout or anatomic imaging sequence, is applied to acquire coil element-specific MRI images. These images define localized sensing volumes associated with individual coil elements and can be acquired either before or after the recording of neuronal firings. In the active strategy, a single-voxel selective excitation pulse sequence, such as single-voxel spectroscopy (SVS), is used to explicitly define a target volume for detecting neuronal firings.

Neuronal firings are recorded using an FID-based pulse sequence, such as *fid* or *SVS* without water suppression, to acquire FID signals from each coil element over multiple repetition times (TRs). The resulting data consist of complex-valued time series sampled at a temporal resolution $\Delta t$. These FID signals are processed using Equations (1) – (4) and the flowchart in Figure 1f to estimate the neuronal magnetic field component $B_{n,z}$ along the main magnetic field ($B_0$) direction of the MRI scanner.

**Figure 1: Conceptual schematic of the qsMRI for detection of neuronal firings.**

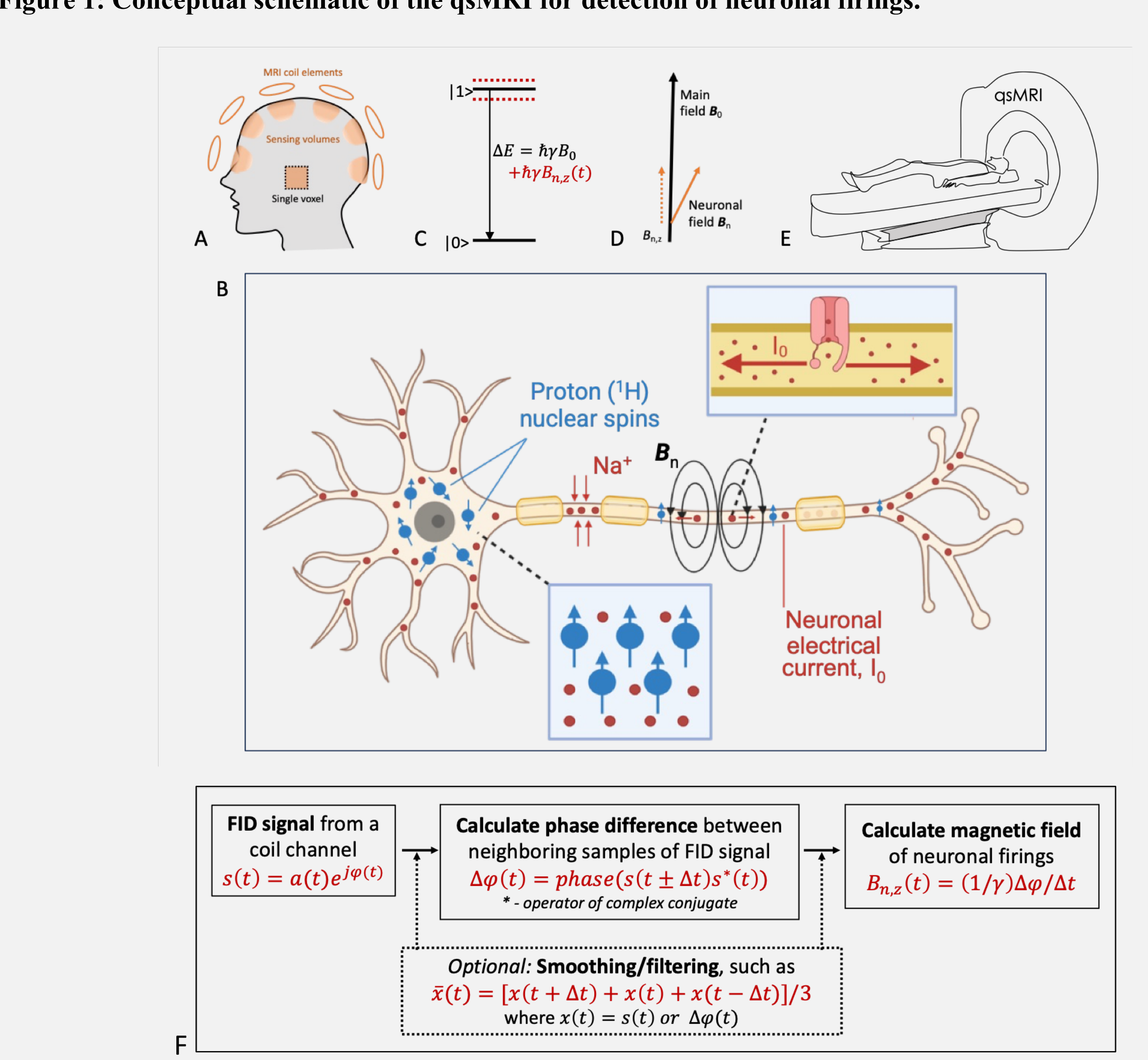

**a,** Sensing volumes (shading areas) of the qsMRI defined either by coil element sensitivity or by single-voxel-select excitation (dashed line). **b,** Endogenous quantum sensors of water proton ($^1$H) nuclear spins (blue dots with arrows) immersed in sodium ionic ($Na^+$) flow (neuronal current) inside a firing neuron and the axon. **c,** Quantum sensing of neuronal magnetic field $B_n(t)$ via the nuclear spins transiting from excited state |1> to ground state |0>. **d,** Modulation of $B_{n,z}$ onto the main magnetic field $B_0$. **e,** Recording of neuronal firings via an MRI system. **f,** Flowchart of the data processing.

$$\varphi(r,t) = \varphi(r,TE) + \gamma \int_{TE}^{t} B_{n,z}(r,\tau)d\tau \quad (1)$$

$$s(t) = \int_{\Delta V} c(r)\rho(r)e^{-t/T_2^*(r)}e^{-j\varphi(r,t)}dr \quad (2)$$

$$\Delta\varphi(t) = phase[s(t+\Delta t)s^*(t)] \quad (3)$$

$$\widetilde{B_{n,z}}(t) \approx (1/\gamma)\Delta\varphi(t)/\Delta t \quad (4)$$

where $\gamma$ is the gyromagnetic ratio of the nucleus. The phase $\varphi$ in Equation (1) is modulated by the time-varying neuronal magnetic field $B_{n,z}$ and by static local field inhomogeneity $\Delta B_0$ (not shown). The modulation is encoded in FID signal $s(t)$ in Equation (2), from which an estimate $\widetilde{B_{n,z}}$ of the neuronal magnetic field $B_{n,z}$ is derived using Equations (3) and (4). The estimate represents a weighted average over the sensing volume $\Delta V$, determined by the magnetization density $\rho(r)$, effective transverse relaxation time constant $T_2^*(r)$, and coil element sensitivity map $c(r)$. The estimate converges to the true field as $\Delta V$ decreases or as the coil sensitivity becomes more spatially localized. The data-processing workflow is summarized in Figure 1f.

**Estimate of neuronal magnetic field inside a firing axon**

Concerns have been raised regarding whether qsMRI can detect neuronal firings given the extremely weak magnetic fields reported in literature. For example, the magnetic field measured at eight axon radii (~0.107 mm) has been estimated to be ~0.2 nT,[11] which would induce an undetectably small phase shift (< 9.75º×10$^{-5}$) for proton nuclear spins sampled at an interval of 0.2 ms. Consistent with this, ex vivo measurements have reported similarly week fields: 2.5 nT from a single worm axon,[12] ~0.06–0.07 nT from a frog nerve fiber,[13] and ~0.2 nT from a crayfish giant axon.[11] Analytical models based on membrane potential waveforms have yielded comparable estimates (~0.6 nT ).[14-16] Moreover, fMRI phantom studies using current-carrying wires detected no measurable phase shifts,[17] and recent in vivo MRI studies in mice and turtle have produced inconsistent results.[18-20] Collectively, these studies confirm that neuronal magnetic fields measured *outside* axons are extremely weak and effectively undetectable with conventional MRI. However, they share a critical limitation: measurements and models focused on the *extracellular* space, where the neuronal magnetic field exhibits a quadrupolar geometry and rapidly decay with distance. In contrast, substantially stronger dipolar magnetic fields exist *inside* firing axons – precisely the regime targeted by qsMRI for direct detection of neuronal firings.

We estimated the neuronal magnetic field inside a firing axon with a typical diameter of 0.1–1.0 µm. During an action potential, the opening of voltage-gated sodium channels on the axonal membrane allows sodium ions ($Na^+$) to enter the axon, forming two oppositely directed dipolar electric currents along the axonal axis. Using a simplified model of a long, straight current-carrying conductor, we estimated the magnetic field generated by a single dipolar current to be 18–182 μT (Table 1). Although modest in absolute magnitude, such fields are sufficient to induce measurable phase shifts in FID signals, corresponding to 8.78º – 88.78º for proton nuclear spins at a temporal sampling internal of 0.2 ms.

**Table 1: Estimate of neuronal magnetic field *inside* a firing axon.**

| | |
|---|---|
| Magnetic field (long-straight line model) | $B_{n,xy} = \frac{\mu_0 I}{2\pi r} = \left(\frac{1}{2}\right)\mu_0\, e\, D\, v\, r$, $0 < r \leq a$ |
| Magnetic permeability | $\mu_0$ = 1.256x10$^{-6}$ Tm/A |
| Elementary charge | *e* = 1.602 x 10$^{-19}$ C |
| Density of charged particles | *D* = 150 mM, for $Na^+$ during firings |
| velocity of action potential | *v* = 20 m/s |
| **Axon radius, *a*** | **Magnetic field, $B_{n,xy}$ at *r* = *a*** |
| 0.1 μm | 18.2 μT |
| 1.0 μm | 181.7 μT |

The encouraging estimates above were derived at the level of a single axon and do not, by themselves, guarantee detectability of firing-induced magnetic fields in FID signals acquired from a macroscopic sensing volume $\Delta V$ containing tens of thousands of neurons.[21] First, the cylindric symmetry of neuronal magnetic field outside an axon is expected to cause phase cancellation in MRI signals, as implied in Equation (2), rendering conventional detection unlikely. However, such cancellation may not occur when microscopic quantum sensors – nuclear spins – reside inside axons, where the local microenvironment is heterogeneous and quantum phase evolution dominates. Second, neuronal magnetic fields $B_{n,z}$ are unlikely to be spatially uniform across volume $\Delta V$. Under this realistic condition, the relationship between phase and magnetic field described by Equations (3) and (4) becomes nontrivial, and the estimated $B_{n,z}$ represents a weighted average over $\Delta V$, potentially affecting both sensitivity and accuracy. These considerations motivate further evaluation of qsMRI performance through numerical simulations and physical modeling.

## Accuracy of qsMRI measurements

Direct assessment of qsMRI accuracy requires a ground-truth reference for neuronal magnetic fields inside axons, which is currently unavailable due to the lack of suitable physiological models. We therefore evaluated accuracy using numerical simulations. Two models of neuronal magnetic fields $B_{n,z}$ were constructed, each consisting of firings with fixed duration but varying amplitude and polarity (Figure 2a): (i) six action potentials (2ms duration for each) and (ii) three postsynaptic potentials (40ms duration for each). Using Equations (1) and (2), we generated FID signals from a uniform sensing volume $\Delta V$ with representative acquisition parameters (sampling interval 0.0977 ms, SNR = 250). Independent Gaussian noise was added to the real and imaginary components of FID signals across the 16 channels. Neuronal magnetic fields were then estimated from the noisy FID signals using Equations (3) and (4) (Figure 2b). Accuracy was quantified by comparing the peak amplitudes of the estimated fields with the corresponding ground-truth values in the simulated inputs (Figure 2a). The mean percent measurement accuracy of qsMRI was found to be 0.17% for the action potentials and 0.09% for the postsynaptic potentials (Figure 2c). Measurement precision was also assessed using the coefficient of variation, defined as $\delta = \mathrm{SD}/\mathrm{Mean}$, yielding $\delta$ = 0.06% for the action potentials and $\delta$ = 0.53% for postsynaptic potentials. Together, these results demonstrate that qsMRI can achieve high accuracy and high precision in neuronal firing measurements under realistic noise conditions.

**Figure 2: Simulation assessment of qsMRI measurement accuracy.**

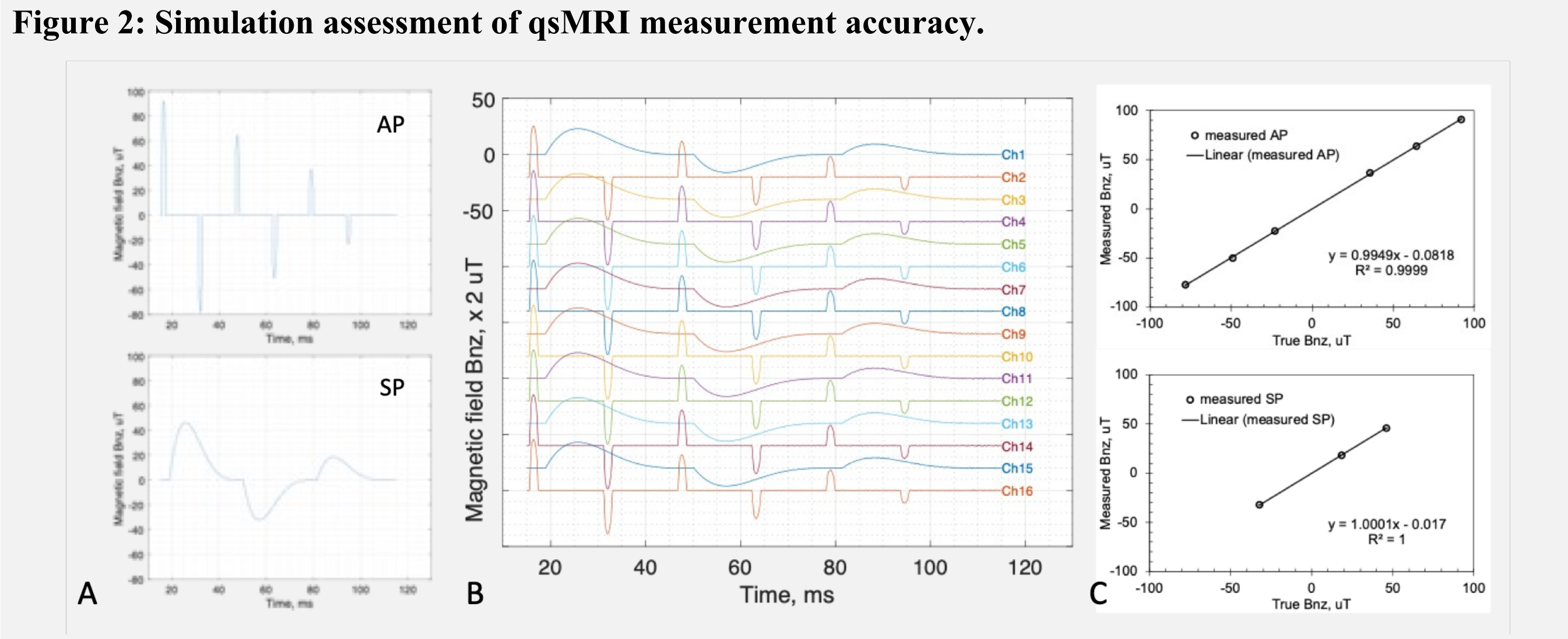

**a,** Numerical models of neuronal magnetic field $B_{n,z}$ of postsynaptic potentials (40ms-long each) and action potentials (2ms-long each). **b,** qsMRI measurements decoded from the noisy FID signals (not shown). **c,** Comparison of the peak values between the qsMRI estimates in **b** and the true values in **a.** The error bars in **c** are smaller than markers and barely visible.

## Definitive localization of neuronal firing sources

Accurate localization of neuronal firing sources remains a major limitation of scalp EEG and MEG, which infer intracranial activity from extracranial recordings by solving an ill-conditioned inverse problem, resulting in substantial spatial uncertainty.[22, 23] In contrast, qsMRI provides an active localization strategy that overcomes this limitation. Here we demonstrate one of two qsMRI localization approaches: exploiting the spatially localized sensitivity of individual coil elements (Figure 1a). In a human subject (Figure 3), qsMRI first acquired anatomical images (Figure 3a) and then used uncombined, coil element-specific images to define the sensing volumes corresponding to the acquired FID signals (Figure 3b). Because each FID signal is intrinsically linked to a known coil sensitivity profile, the neuronal firing sources can be localized definitively without solving an inverse problem.

**Figure 3: Definitive localization of neuronal firing sources via qsMRI images.**

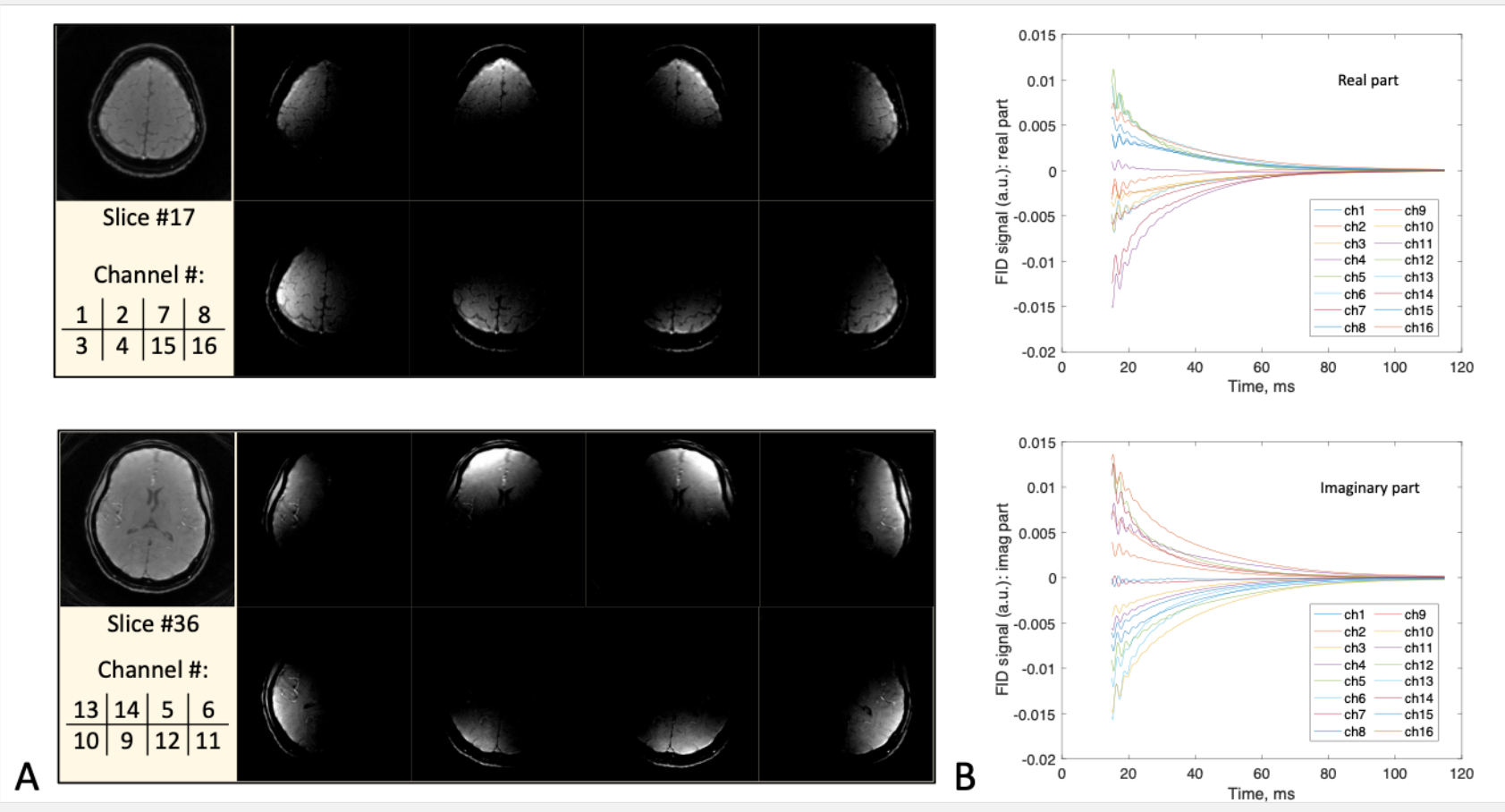

**a,** MRI images of a human subject's brain at 3T: combined from all the 16 head channels (left) and uncombined from the individual channels (right) which define qsMRI sensing volumes. The 16-channel coil elements were configured in two rows, covering the top (slice #17) and mid (slice #36) portions of the brain. **b,** Acquired FID signals during one TR from the 16 coil channels, having SNR = 675 (Ch8) – 2551 (Ch1). These coil element-specific MRI images definitively determines the location of neuronal firings recorded in the FID signals.

## Baseline study on static phantoms

Because qsMRI is designed to detect signals arising from dynamic neuronal electrical activity, no firing-related signal should be present under static conditions. To test this baseline expectation, we performed qsMRI experiments on a non-firing phantom using a clinical 3T MRI system (Prisma Fit, Siemens) equipped with a 20-channel Head/Neck array coil. The phantom was a standard Siemens-provided regular model consisting of a 2000-mL plastic bottle filled with saline solution (0.5% w/w NaCl). All 16 head coil channels were activated to ensure full spatial coverage. Figure 4a shows images reconstructed from combined and individual channels, and the FID signals acquired from each channel are shown in Figures 4b–d. The derived qsMRI recordings exhibit flat profiles (Figure 4e), confirming the absence of detectable magnetic field perturbations under non-firing conditions.

## Neuronal firing study in humans

All human studies were approved by the author's Institutional Review Board (IRB), and written informed consent was obtained from all participants.

### *Detection of neuronal firings*

Neuronal firing activity persists even at resting state, in the absence of intentional external stimulation, as indirectly inferred from functional MRI.[24] In contrast, qsMRI is designed to *directly* detect such electrical activity. Figure 5 presents representative qsMRI recordings acquired from a healthy young participant (25-year-old male) using a clinical 3T MRI scanner (Prisma Fit, Siemens) with a 20-channel Head/Neck array coil. Figures 5a–c show the magnitude, phase, and unwrapped phase of the acquired FID signals, with a consistent color map across panels. Massive signal peaks were observed during the initial ~80 ms acquisition window across multiple channels (Figure 5d), including channels over the prefrontal (Ch7, Ch14) and occipital (Ch15 and Ch16) regions (Figure 5e). Enlarged views of representative peaks (Figure 5f) show an average duration of 1.4±0.3 ms (7.2±1.5 data points), consistent with the temporal scale of neuronal action potentials (~2 ms). In addition, a burst of firing activity was detected in Ch5 (Figure 5g). These features cannot be explained by phase wraparound artifacts (Figure 5d), which typically affect only a few adjacent data points. Although random noise may occasionally generate transient fluctuations, such noise-induced features are distinguishable from firing-related peaks by their association with low SNR in corresponding segments of FID signals, as illustrated in regions on the top-right corner and far-right side in Figure 5d, where FID signal amplitude is diminished (Figure 5a). Finally, the peak amplitudes in Figure 5d are well contained within ±100 μT, consistent with the order-of-magnitude estimates summarized in Table 1.

**Figure 4: Phantom study without neuronal firings.**

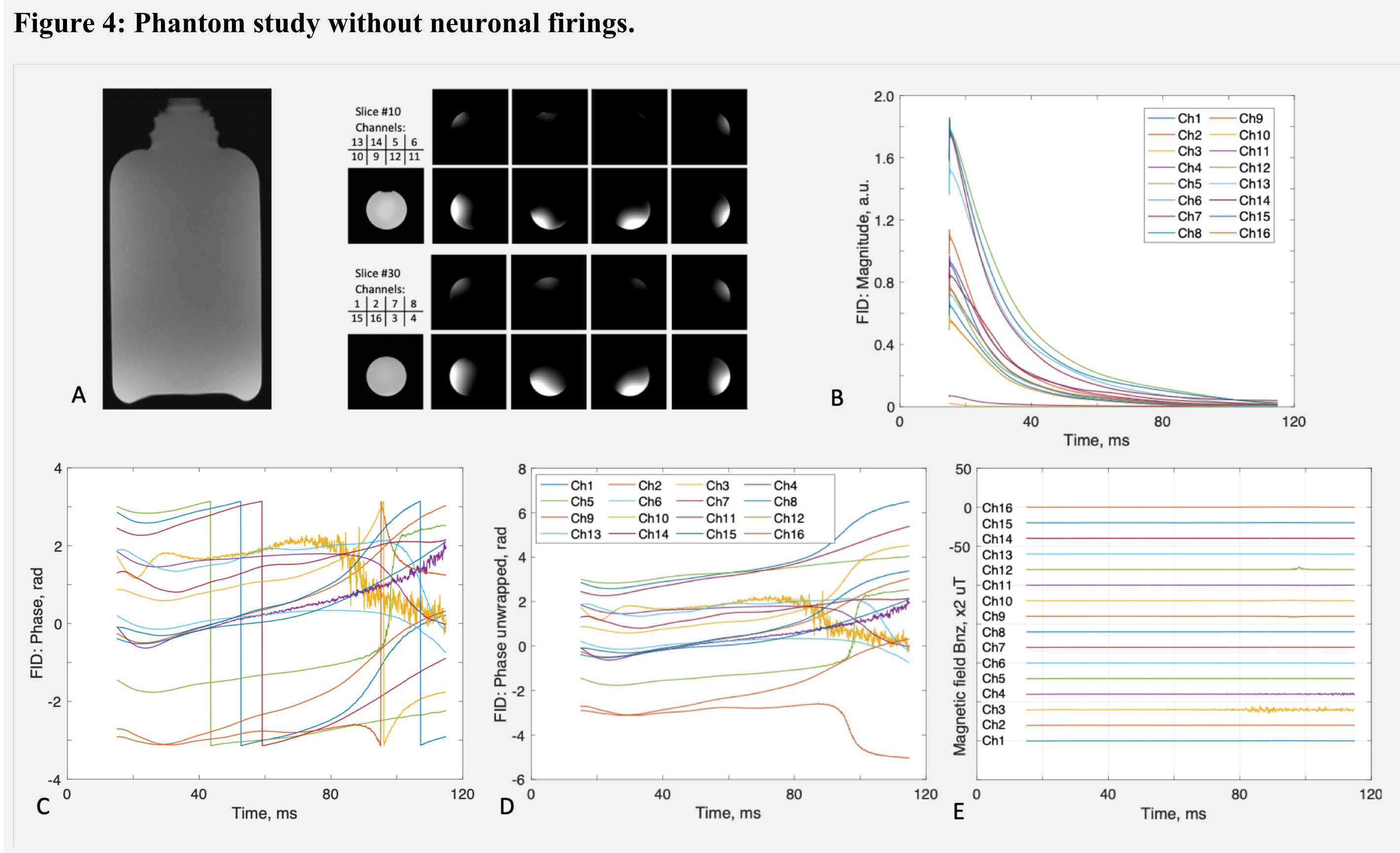

**a,** MRI images of the phantom – a plastic bottle (2000 mL) filled with saline water (0.5% ww in NaCl) – one big sagittal slice and two small transverse slices. The uncombined transverse slice images (displayed in the same window/level) define the sensing volume at individual channels. **b-d,** FID signals (magnitude, phase, and unwrapped phase) at each of the 16 channels. **e,** The derived qsMRI recordings (neuronal firing magnetic field) calculated from the FIDs in b-d.

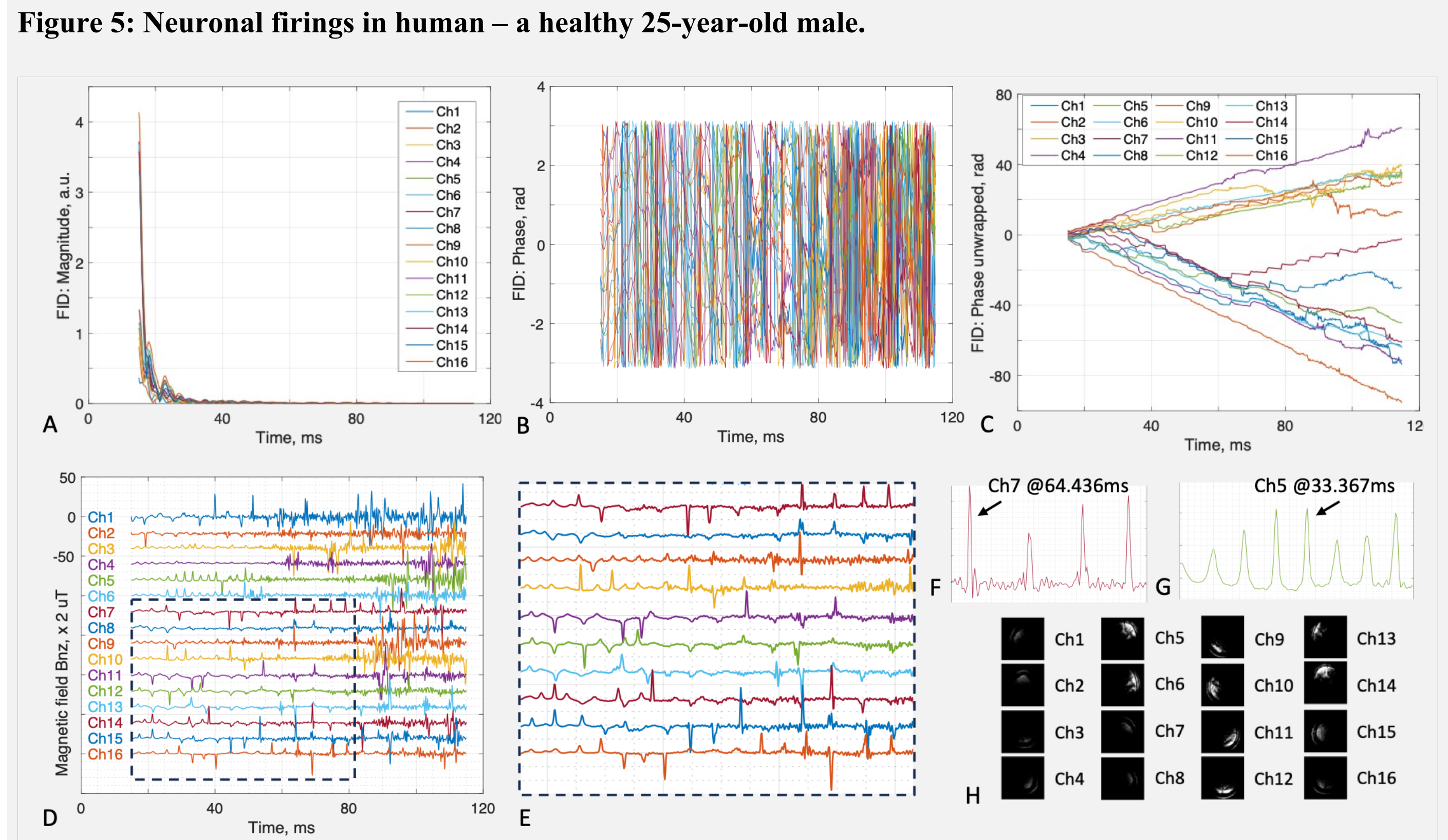

**Figure 5: Neuronal firings in human – a healthy 25-year-old male.**

**a–c,** FID signals acquired at the head channels 1–16 during one readout: magnitude, phase, and unwrapped phase. **d,** qsMRI recordings calculated from the FIDs in a–c, showing massive firing peaks, both positive and negative. In contrast, random noise field appears on the far-right side and top-right conner. **e,** Enlarged view at the dashed box region in d. **f,** Zoomed-in view of singlet peaks at Ch7 and time 64.426ms in d. **g,** Zoomed-in view of burst firings at Ch5 with the mid-right peak at time 33.367 ms in d. **h,** MRI images at individual channels, showing the locations of neuronal firing sources. The line color map is consistent across a–g.

*Neuronal firings across ages at resting state*

To observe whether qsMRI can detect neuronal firings at the individual-subject level across different ages and sexes, we studied 25 cognitively normal subjects at resting state (awake or asleep) in the early morning (~7:00 AM). The subjects were 27–84 years old, or 58.8±18.2 years in mean and standard deviation (SD), including 9 males and 16 females. Resting-state FID signals were acquired ~30 min after the start of the MRI session using a product sequence *fid* on a 3T MRI scanner (Prisma Fit, Siemens) with a standard 20-channel Head/Neck array coil. The FID readout time was 819.2 ms with a sampling interval 0.2 ms, TE/TR = 0.2/1500ms, 64 repetitions, and manual $B_0$ shimming. Neuronal firing rate $R_f$, defined as the number of firing peaks per second (Hz), was quantified by counting the number of firing peaks $N_{pk}$ within the initial time window $T_{pk}$ = 500ms of high SNR at each FID signal, i.e., $R_f = N_{pk} / T_{pk}$, using an automatic peak-detection algorithm.[25] The mean and SD of firing rate were computed across the 64 repeated FID acquisitions at each channel. Whole-brain firing rate for each subject was then attained by averaging the channel-wise means and SDs across the 16 head channels. Figure 6a presents the resulting firing rates for individual subjects, and the corresponding percent variability (SD/mean) is shown in Figure 6b. These relatively small intra-subject variations, either absolute or percent (22.3±7.7 %), indicate that qsMRI is capable of reliably resolving whole- brain firing rates at the individual-subject level.

**Figure 6: Neuronal firings across participants (ages) at resting state.**

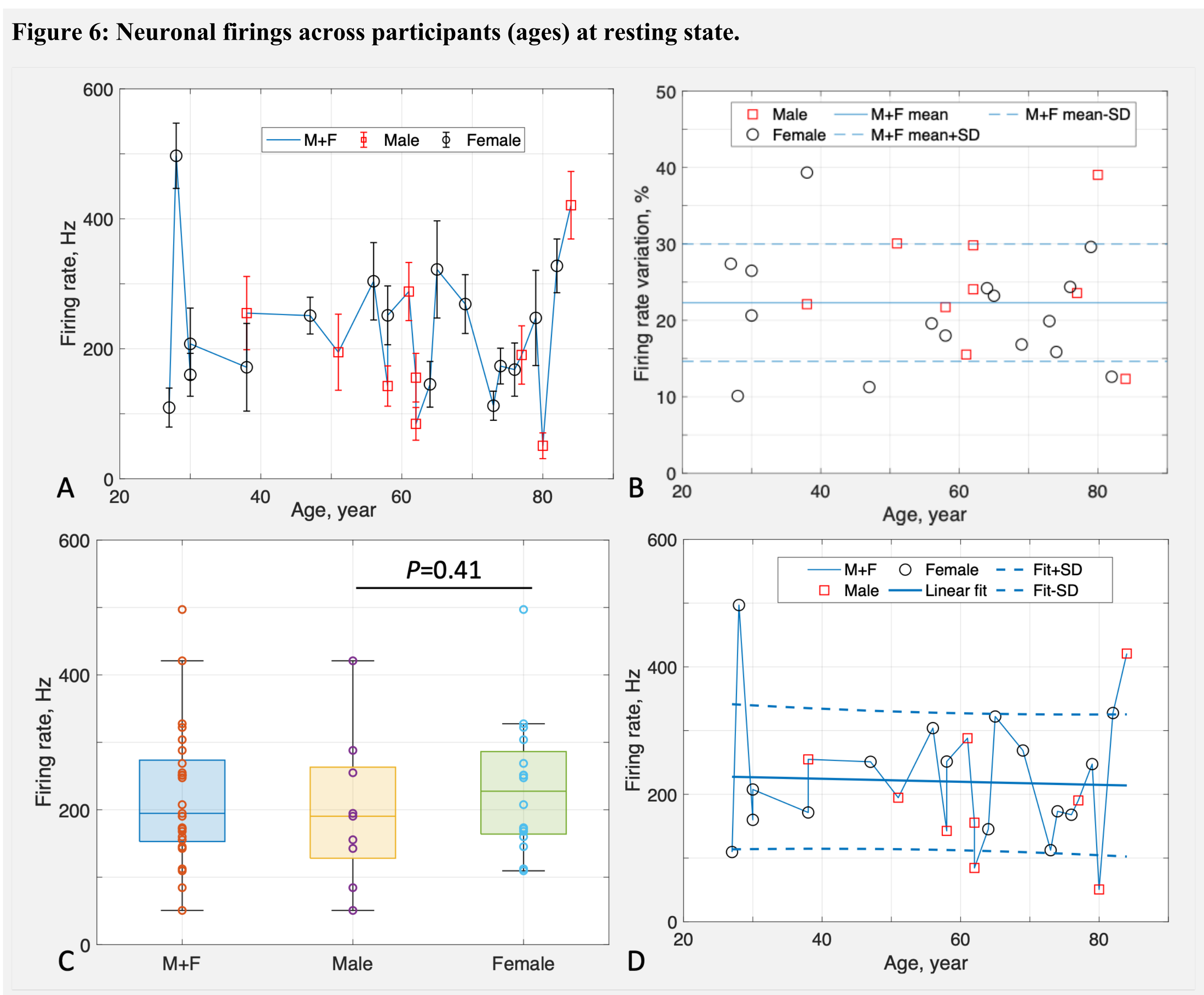

**a,** Neuronal firing rate in whole brain (averaged over all the 16 head channels) from a group of 25 healthy subjects (9 males and 16 females), showing the potential to differ firing rate at an individual-subject level. **b,** Percent variation (SD/mean) in firing rate for each subject, with an overall percent variation of 22.3±7.7 %. **c,** Grouped distributions of the firing rate, showing no statistically significant difference between the subgroups: $P = 0.41$ for females vs. males, 0.70 for females vs. combined (M+F), and 0.57 for males vs. combined (M+F). **d,** Linear regression between firing rate and age over all the subjects, with no correlation ($R^2 = 0.0018$).

To explore sex-related difference, we divided the participants into male and female subgroups, and group comparisons were performed using the non-parametric Mann-Whitney U test, with statistical significance defined as $P < 0.05$. As shown in the scatter box-and-whisker plot (median, 1st and 3rd quartiles) in Figure 6c, no statistically significant differences were observed between the groups: $P = 0.41$ for females versus males, $P = 0.70$ for females versus the combined cohort (males + females), and $P = 0.57$ for males versus M+F. Although not statistically significant in the current cohort, the female subgroup exhibited a trend toward separation from the male subgroup ($P = 0.41$), which may warrant further investigation in studies with larger sample size. Notably, the female subgroup showed two apparent clusters of data points above and below the median firing rate (227.5 Hz), suggesting potential heterogeneity in firing patterns within this subgroup.

To explore age-related effects, we performed a linear regression analysis between whole-brain firing rate and age across all the subjects (Figure 6d). No significant association was observed, with a coefficient of determination of $R^2 = 0.0018$, indicating negligible correlation between firing rate and age in this cohort.

*Neuronal firings under a task of finger tapping*

This controlled-stimulation study provides an initial validation of the qsMRI technique. Finger tapping is known to elicit increased neuronal firings in the motor cortex compared with the absence of a task. We hypothesized that qsMRI could detect task-induced changes in firing rate at the individual-subject level. A cohort of 25 healthy participants – the same subjects shown in Figure 6 – performed a finger-tapping task using the index finger of their dominant hand (Fig. 7a). Participants were instructed to tap at approximately 1.0 Hz for 1.5 min. Two additional functional states were acquired as references. First, a no-tapping state was measured approximately 15 s after the completion of the tapping task, allowing subjects to return to a calm state under operator monitor; this acquisition used the same $B_0$ shimming and field of view (FOV) as the tapping condition. Second, a resting state scan was acquired about 30 min after the no-tapping condition, with $B_0$ re-shimming. This resting-state acquisition was the same one as shown in Figure 6.

The experimental setup is illustrated in Figure 7a and includes: (i) the task paradigm (finger tapping, no tapping, and resting); (ii) the element configuration of the 20-channel Head/Neck array coil, comprising anterior elements (H1, H3, N1) and posterior elements (H2, H4, N2); (iii) a schematic brain illustration indicating the motor cortex; and (iv) partial MRI images reconstructed using a sum-of-squares method from motor-related coil elements only (H21–H24).

Representative qsMRI recordings on a participant (58-years-old male) under the three task conditions are shown in Figure 7b, with three consecutive acquisitions for each condition. Channels 1, 2, 17, and 18 correspond to neck coil elements (N12, N11, N22, and N21) and are dominated by random noise impact.

We next examined firing rates across participants (ages) and between sex subgroups. The firing rate was calculated within the initial 500ms time window. Substantial inter-individual variability in firing rate was observed across ages (Figure 7c), particularly among late middle-aged and early older adults (55–75 years).

The grouped plots in Figure 7d further illustrate these variations, showing firing-rate distribution under the three conditions: median [Q1–Q3] = 133.9 [78.5–148.4] Hz for males; 162.7 [88.7–260.1] Hz for females; and 139.4 [90.7–208.0] Hz for the combined cohort. However, the three conditions did not show statistically significant

**Figure 7: Neuronal firings under a task of finger tapping.**

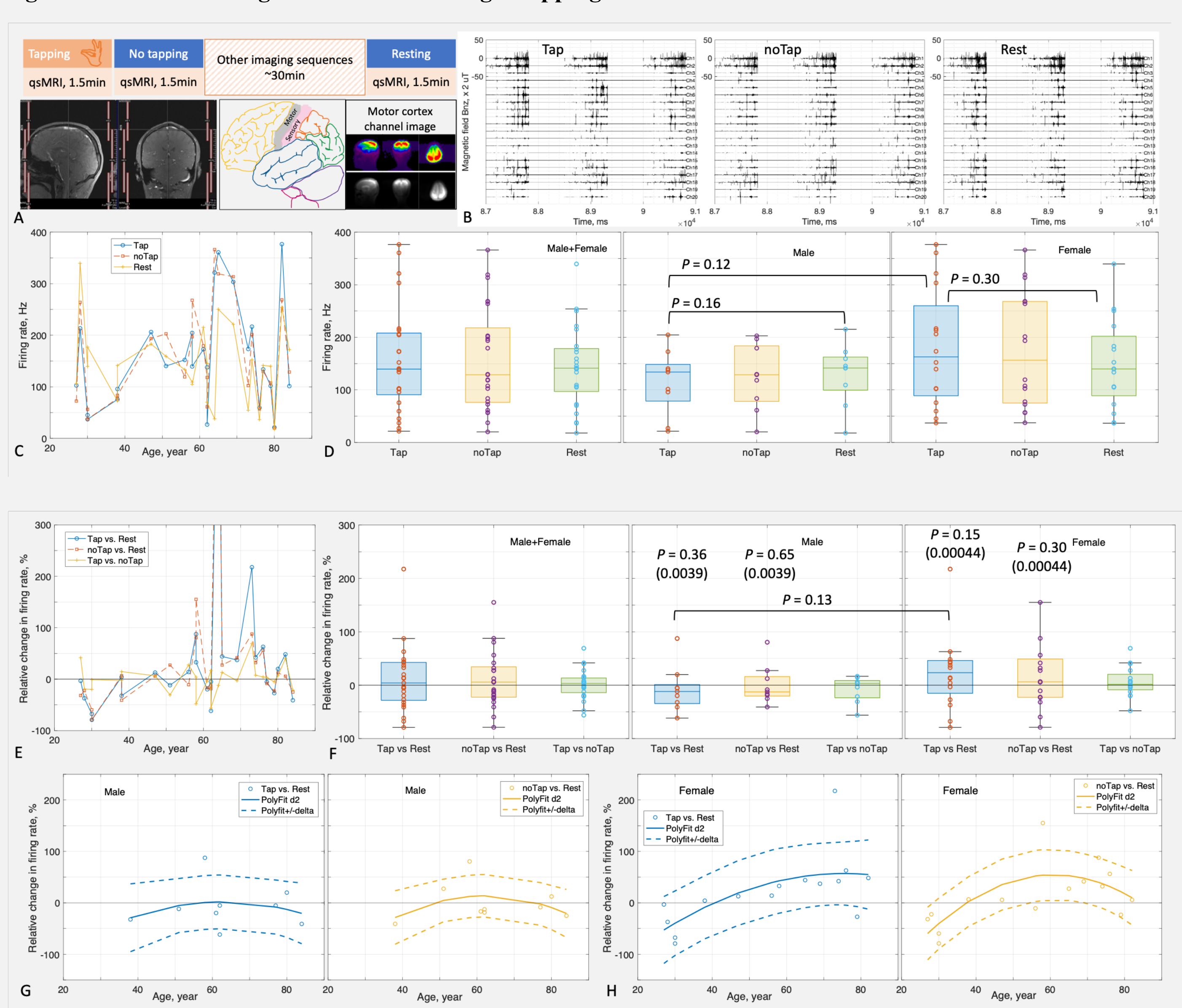

**a,** Experimental set-up: a task paradigm including finger tapping, no tapping, and resting; a coil element configuration of the 20-channel Head/Neck array coil, including anterior element groups (H1, H3, N1) and posterior element groups (H2, H4, N2); a brain scratch indicating motor cortex location; MRI images reconstructed from the motor cortex-related channels (H23, H24, H21, H22). **b,** Representative qsMRI recordings on a participant (58-year-old male). **c,** Neuronal firing rate across participants (ages). **d,** Distributions of firing rate for the male, female, and combined (male + female) groups. **e,** Relative change of firing rate across ages, with two extreme outliers (outside the figure) from one female subject aged 64 years: 750.2% in Tap vs. Rest, and 868.0% in noTap vs. Rest. The large fluctuating change (>25%) occurred at ages 58–75 years. **f,** Distributions of relative firing-rate for the male, female, and combined groups, showing a trend of difference ($P$ = 0.13) between the male and female subgroups in Tap vs. Rest. **g & h,** Regression analysis (polynomial fitting, 2nd degree) of the relative firing-rate in the male and female subgroups, showing the older ages (>58 years) in the female subgroup have a higher relative firing-rate during tapping than in the male subgroup, with $R^2$ = 0.41 (Tap vs. Rest) and 0.49 (noTap vs. Rest) for the females, while $R^2$ = 0.07 and 0.20 for the males, respectively. Note: the two outliers are not shown in the female subgroup in f and not included in the regression in h.

differences in firing rate, either within a group under the Wilcoxon Signed-Rank test between tapping vs. resting and no-tapping vs. resting conditions: $P = 0.16$ and 0.57 for the males, $P = 0.30$ and 0.44 for the females, and $P = 0.72$ and 0.70 for the combined cohort; or between the female subgroup ($N = 16$) and male subgroup ($N = 9$) under the Mann-Whitney U test: $P = 0.12$, 0.44, and 0.84 for the tapping, no-tapping, and resting conditions, respectively. Nevertheless, we observed a trend toward statistical significance within the males for tapping against resting ($P = 0.16$) and between the females and males for tapping ($P = 0.12$), as summarized in Table 2.

**Table 2: *P*-values of firing rates under three task conditions.**

| Firing rate | | | | |
|---|---|---|---|---|
| | Tap vs Rest | noTap vs Rest | Tap vs noTap | Test |
| Female (*N* = 16) | 0.3011 | 0.438 | 0.6417 | WSR* |
| Male (*N* = 9) | **0.1641** | 0.5703 | 0.6523 | WSR |
| Combined (F+M) | 0.7164 | 0.6964 | 0.8824 | WSR |
| | Tap | noTap | Rest | |
| Female vs Male | **0.1195** | 0.4447 | 0.8429 | MWU* |
| Relative firing-rate change Δ and absolute value \|Δ\| | | | | |
| | Tap - Rest | noTap - Rest | Tap - noTap | |
| Female | **0.1477** | 0.3011 | 0.5014 | WSR |
| Female \|Δ\| | **0.00043778** | **0.00043778** | **0.00043778** | WSR |
| Male | 0.3594 | 0.6523 | 0.7344 | WSR |
| Male \|Δ\| | **0.0039** | **0.0039** | **0.0039** | WSR |
| Combined (F+M) | 0.4118 | 0.5272 | 0.7164 | WSR |
| Combined \|Δ\| | **0.00001229** | **0.00001229** | **0.00001229** | WSR |
| Female vs Male | 0.1335 | 0.4117 | 0.4117 | MWU |
| F vs M \|Δ\| | 0.2696 | 0.2949 | 0.9774 | MWU |

**Note: WSR – Wilcoxon Singed Rank test; MWU – Mann-Whitney U test.*

We then turned to the relative changes in firing rate under the tapping and no-tapping conditions compared with the resting state (Figures 7e and 7f). Two extreme outliers from a single female participant (64 years old) were out of display range: 750.2% for tapping vs. resting and 868.0% for no-tapping vs. resting. The large fluctuations in relative change (>25%) were observed predominantly in participants aged 58–75 years. The distributions of relative firing-rate changes are also shown for males, females, and the combined groups. A trend toward a sex difference was observed for tapping vs. resting, although it did not reach statistical significance (Mann-Whitney U test, $P = 0.13$). No significant differences were detected between female and male subgroups for no-tapping vs. resting ($P = 0.41$) or for tapping vs. no-tapping ($P = 0.41$). We further examined the relative firing-rate changes (as well as their absolute values) within the males and females and found a trend toward statistical significance only in female's tapping vs. resting ($P = 0.15$). However, when examining the absolute value (or magnitude) of changes, we found all the relative firing-rate changes statistically significant: $P = 0.00044$ for the females in tapping (no-tapping) vs. resting and $P = 0.0039$ for the males as well (Table 2).

Finally, we performed age-based regression analysis of the relative firing-rate changes separately for male and female subgroups. The optimal polynomial fit was achieved with a second-degree model using the *polyfit()* function in MATLAB, as shown in Figures 7g and 7h. The regression results indicate that, at older ages (>58 years), the female subgroup exhibits larger relative firing-rate increases under task condition compared with the male subgroup. For females, the coefficients of determination were $R^2$ = 0.41 for tapping vs. resting and $R^2$ = 0.49 for no-tapping vs. resting, better than the corresponding values for the males, $R^2$ = 0.07 and $R^2$ = 0.20, respectively.

*Neuronal firings in a healthy participant with a history of epilepsy*

This resting-state case study arises from an accidental finding in an otherwise healthy participant (33-year-old female) with a self-reported history of epilepsy. Even when clinically well controlled, epilepsy can be associated with persistent abnormal neuronal electrical activity that is not evident daily functioning. qsMRI may therefore offer the capability to detect such asymptomatic but aberrant neuronal firing patterns. Figure 8a presents representative qsMRI recordings from FID #7 and FID #17 selected from a total of 64 FID acquisitions. Data were acquired using channels Ch1–Ch20, corresponding to head coil elements H11–H44. A markedly increased number of neuronal firing peaks (arrows, zoomed views) was observed in Ch15 and Ch16 across all 64 FID acquisitions, yielding average firing rates of 357.0 ± 71.8 Hz at Ch15 and 408.4 ± 65.2 Hz at Ch16 (Figure 8d). Temporal dynamics of firing rate variation relative to the averaged firing rate across TRs is presented in Figure 8e. As a reference, an age- and sex-matched healthy participant (33-year-old female) without a history of epilepsy was also examined (Figure 8b). Substantially lower firing activity was observed, with average firing rates of 100.8 ± 35.4 Hz at Ch15 and 120 ± 68.9 Hz at Ch16 (Figure 8f), with the associated temporal dynamics shown in Figure 8g.

We next localized the neuronal firing sources associated with channels Ch15 and Ch16 using partial MRI reconstructions from the epilepsy-history participant (Figure 8c). The partial MRI images (color scale; R denotes the right side of the participant) were reconstructed using only channels Ch15 and Ch16, corresponding coil elements H34 and H33, respectively, via the standard sum-of-squares method. The imaging raw data were acquired during $B_0$ shimming immediately prior to the qsMRI FID acquisition. For anatomical reference, full MRI images (grayscale) were reconstructed using all the 16 head channels. Based on these partial MRI reconstructions, channels Ch15 and Ch16 predominately received FID signals from the left frontotemporal lobe. The counter-lateral side (right frontotemporal lobe) correspond to coil elements H32 and H31 or channels Ch5 and Ch6 in Figures 8a,b.

**Figure 8: Neuronal firings in a healthy participant with a history of epilepsy.**

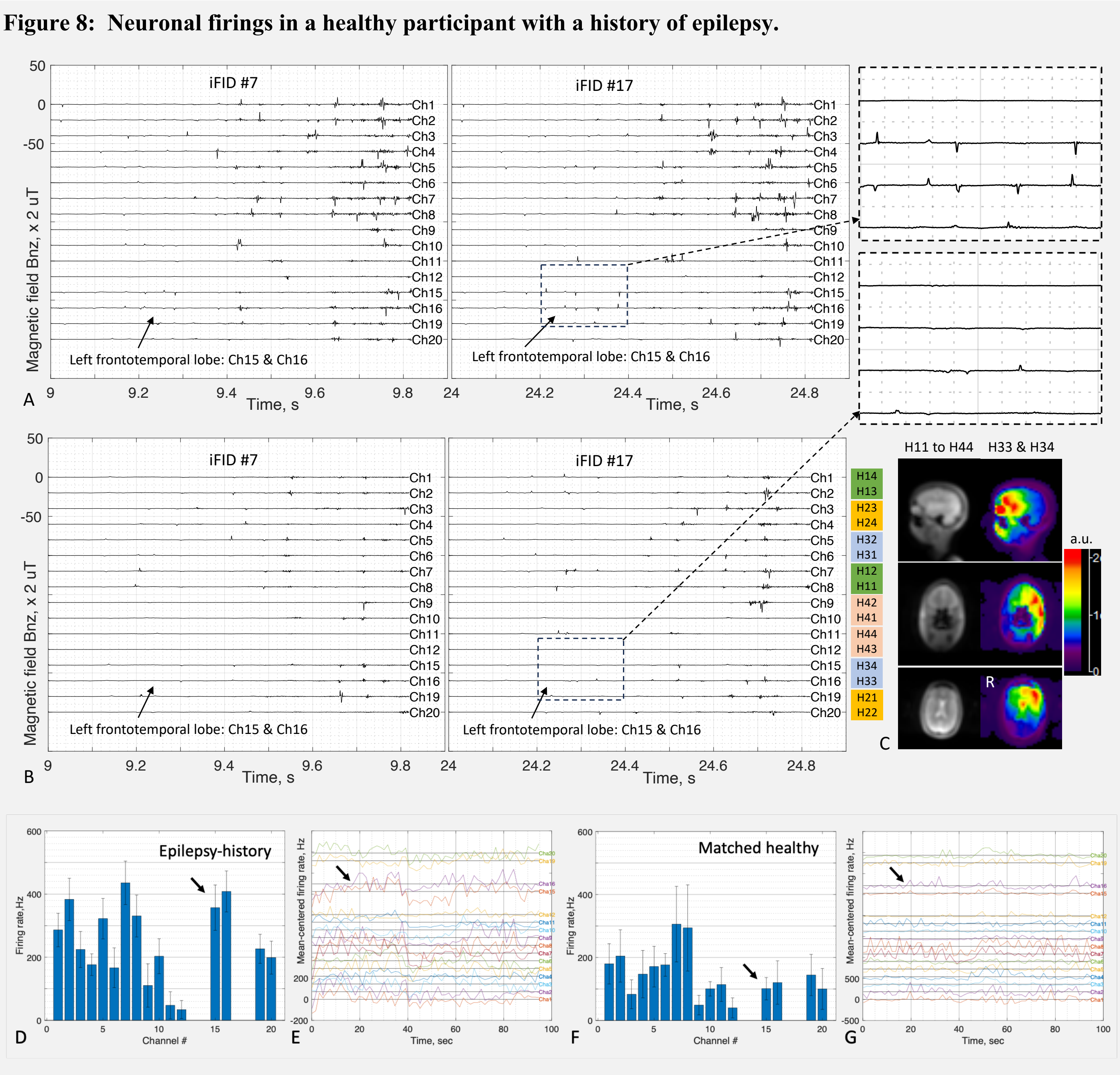

**a,** Representative resting-state qsMRI recordings at FIDs #7 and #17 out of the 64 FIDs on a healthy participant with a history of epilepsy (33-year-old female). The channel indices 1–20 and corresponding coil elements H11–H44 are shown in color-coded below. **b,** Reference case of an age-sex-matched healthy participant (33-year-old female) without epilepsy history. **c,** MRI images of the epilepsy-resolved participant, reconstructed both with all the 16 head elements (gray scale) and with only two head elements (H33 and H34, color scale, R for right side of the person) to confirm Channels 15 and 16 receiving FID signals from the left frontotemporal lobe where neuronal firings were observed in **a** but not in **b**, as zoomed view in the dashed-line regions. The contralateral side (right frontotemporal lobe) is located at elements H31 (Ch6) and H32 (Ch5). **d & e,** Mean (±SD) firing rates over a full recording period of 1.5 min at the 16 head channels and their temporal dynamics across all the 64 TR-recordings for the epilepsy-history subject, and **f & g,** for the age-sex-matched healthy subject.

## Open validation

We note that the results presented in Table 1 and Figures 2–8 constitute a partial, rather than complete, validation of qsMRI measurements of neuronal firings. Specifically, we have not included experiments using a ground-truth model, which remains technically challenging to construct and is not currently available in our laboratories. To facilitate further validation, inspired by the Cogitate Consortium, et al,[26] we openly invite independent verification by researchers with access to suitable ground-truth models. To that end, we provide the following experimental protocols to enable ground-truth-based validation of qsMRI and the Supplementary Information including computation codes and intermediate measurement values for confirmation during the validations.

*A suitable ground-truth model*: It can generate action potentials and/or postsynaptic potentials with controlled waveform, duration, and firing frequencies, and transmit these signals along myelinated axons with well-defined nodes of Ranvier. At each node, paired dipolar electric currents are produced within the axon, flowing in opposite directions – an essential feature of neuronal signal propagation. These paired currents generate neuronal magnetic field both inside and outside the axon: a dominant dipolar field within the axon and a much weaker quadrupolar magnetic field outside the axon. qsMRI is sensitive specifically to the intra-axon dipolar magnetic field, particularly its component $B_{n,z}$ aligned with the main static field $B_0$. Importantly, the microscopic ionic environment within the axon must be significantly inhomogeneous. Sufficient uniform would lead to cancellation of the magnetic-field effects on proton nuclear spins, rendering the neuronal magnet field undetectable by qsMRI.

*FID signal acquisition*: MRI experiments should be performed on a scanner operating at 3T or higher to ensure sufficient SNR in the FID signals. The receiver coil may be a multi-channel array to enable localization of firing sources, or a single-channel volume coil combined with selective RF excitation for targeting a specific region of the model. Manual $B_0$ field shimming should be performed to achieve a water linewidth below 30 Hz. Data acquisition should use either *fid* or single-voxel spectroscopy (*SVS*) pulse sequence without water suppression. FID sampling interval (including oversampling) must be ≤ 0.1 ms for an 819.2ms-long readout, and at least 64 FID acquisitions should be collected.

*Firing source localization*: when an *fid* sequence is used in combination with a multi-channel array coil, coil-specific MRI images are required for localization of neuronal firing sources. These images may be obtained through separate anatomical imaging acquisitions (e.g., MPRAGE, FLAIR, or $T_2$-weighted imaging) or by reconstructing images from the $B_0$-shimming data acquired immediately prior to the FID measurements. Image reconstruction may be performed either inline or offline, as demonstrated in Figure 8c.

*Data processing:* Follow the flowchart in Figure 1f. Specifically, (i) Preprocessing. FID signals should first be visually inspected to assess the impact of random noise and to determine an appropriate analysis time window (e.g., 300 or 500 ms) for firing rate calculation. Frequency spectra should also be examined for off-resonance water peak shifts and fat peak interference. If necessary, water peak should be corrected for off-resonance shift and fat peak removed. When random noise is substantial, FID signals may be down-sampled from 0.1 ms to 0.2 or 0.3 ms by averaging two or three neighboring data points. (ii) Neuronal magnetic field estimation. The neuronal magnetic field component $B_{n,z}$ should then be calculated from the (optionally down-sampled) FID signals using Equations (3)–(4). (iii) Firing-rate calculation. neuronal firing rates are computed by counting firing peaks within the predetermined time window and normalizing by the window duration.

## DISCUSSION

The qsMRI technique presented here is the first of its kind. Its underlying concept is adapted from the broader framework of quantum sensing, with the goal of enabling quantum-sensing measurements on MRI platforms at room temperature in living humans. qsMRI leverages intrinsic nuclear spins within the human body – most notably proton ($^1$H) nuclear spins in water molecules – as quantum sensors, eliminating the need for externally administrated labeling agents. In conventional clinical MRI, nuclear spins are primarily exploited through their relaxation properties ($T_1$ and $T_2$) to generate tissue contrast. qsMRI instead repurposes these nuclear spins as quantum sensors capable of detecting fluctuations in local electrical and magnetic fields via their phase perturbation.

The physics behind qsMRI detection of neuronal firings is solid. Although neuronal magnetic fields are extremely weak outside axons (on the order of ~0.2 nT) due to their quadrupolar nature, the intra-axonal magnetic field can be sufficiently strong (~100 μT) because of the dipolar field generated by paired currents at the nodes of Ranvier. This intra-axonal dipolar field is sufficient to induce measurable phase shifts in quantum sensors during FID acquisition. Moreover, qsMRI measures signals from an essemble of microscopic quantum sensors distributed over a macroscopic tissue volume, rather than from a single sensor, thereby achieving sufficient SNR in acquired FID signals. This also helps relax the requirement for high main magnetic field $B_0$.

The qsMRI technique currently supports two approaches for localizing neuronal firing sources: coil-element-specific sensing volumes (passive mode) and selective RF excitation (active mode). Both modes offer opportunities for further enhancement. For example, localization in the passive mode can be improved by increasing the number of receiver coil elements to 32, 64, or more, while in the active mode, selective RF excitation can be extended to multiple voxels using multiband excitation strategies.[27]

The major advantages of the qsMRI technique include: (i) direct detection of neuronal magnetic fields during firing using in situ quantum sensors (intrinsic nuclear spins), (ii) fully non-invasive implementation on existing clinical MRI scanners at room temperature, and (iii) elimination of the need for exogenous contrast agents.

Nevertheless, we acknowledge several limitations of the present study. First, the results reported here are insufficient to fully validate the qsMRI technique due to the absence of experiments using a ground-truth model. Second, FID acquisitions are separated by repetition times (TRs), resulting in segmented rather than continuous recordings of neuronal firings; however, this gap can be minimized in the future implementations through optimized pulse-sequence design. Third, the spatial resolution for localizing firing sources with qsMRI is currently at the mesoscale and therefore coarser than conventional anatomical MRI, although higher spatial resolution may be achievable with smaller coil elements or selective RF excitations strategies.

While we call for open, community-based validation of qsMRI, we also anticipate paradigm-shifting applications in neuroscience research, including the investigation of neuronal firing dynamics within cognitive brain circuits, as well as transformative clinical applications in neurology (e.g., epilepsy and brain tumors) and psychiatry (e.g., depression and bipolar disorder).

In summary, quantum sensing may be translatable to existing clinical MRI platforms operating at room temperature in humans. Pending rigorous independent validation, the qsMRI approach described here could establish a new noninvasive framework for the direct detection of neuronal firing without exogenous contrast agents, with broad implications for neuroscience research and potential clinical translation in neurology and psychiatry.

## METHODS

### *MRI quantum sensing of neuronal firings*

In qsMRI, endogenous proton ($^{1}$H) nuclear spins in water molecules are used as quantum sensors because of their high signal-to-noise ratio. Other MRI-visible nuclei, such as sodium ($^{23}$Na), are also, in principle, applicable. When neurons generate an action potential and propagate it along myelinated axons, sodium ions ($Na^{+}$) from the surrounding extracellular space enter the axons at nodes of Ranvier, giving rise to paired electric currents flowing in opposite directions (parallel and antiparallel to axonal axis). These paired neuronal currents, $\boldsymbol{I_n}$, produce a local magnetic field, $\boldsymbol{B_n}$, characterized by a strong dipolar component within the axon and a much weaker quadrupolar component outside the axon. The resulting neuronal magnetic field can substantially perturb the resonance frequency of nuclear-spin quantum sensors located inside the axon, while its effect on spins outside the axon is negligible. Quantum sensing occurs in the presence of the time-varying neuronal magnetic field component $B_{n,z}(t)$, aligned with the main static magnetic field $\boldsymbol{B}_0$, during the transition of nuclear spins from the excited state |1> to the ground state |0>. The corresponding energy difference $\Delta E$ is modulated by the neuronal magnetic field superimposed on $B_0$. These physical processes can be summarized using the following mathematical and MRI-based descriptions.

The neuronal magnetic field $\boldsymbol{B}_n(r,t)$, induced by neuronal electric currents $\{\boldsymbol{I}_{\mathrm{n}}(r,t), r \in \Delta V\}$, is evaluated at spatial location $r$ and time $t$. Nuclear spins accumulate an additional phase $\varphi(r,t)$, as defined in Equation (1), which is encoded in the FID signal $s(t)$ in Equation (2). Within a quantum-sensing volume ΔV, FID signal $s(t)$ is determined by coil sensitivity $c(r)$, magnetization density $\rho(r)$, and effective transverse relaxation time constant $T_2^*(r)$. To estimate neuronal magnetic field component along the main field direction, $B_{n,z}$, the phase difference of FID signal between successive sampling points separated by a time interval $\Delta t$ is calculated using Equation (3) and subsequently converted into $B_{n,z}$ using Equation (4).

<u>*Simulation evaluation of qsMRI measurement accuracy*</u>

First, we created typical waveforms for action potential (AP) and postsynaptic potential (PP), respectively. The AP waveform was modeled using a Gaussian function, $f_{AP}(t) = exp(-((t-t_0)/2\sigma)^2)$, $0 \leq t < \tau$, with a duration $\tau$ = 2 ms, center time $t_0 = \tau/2$, and $\sigma$ set to match the waveform duration. The PP waveform was modeled as a weighted sine function, $f_{PP}(t) = (1-t)^3 Sin(\pi t)$, $0 \leq t < \tau$, with a duration $\tau$ = 40 ms. We then defined 16 channels, divided into two interleaved groups: one group representing AP channels and the other representing PP channels. Each AP and PP channel was constructed independently. In each AP channels, a chain of six equally spaced AP waveforms was placed, with the first waveform having an amplitude of 92 Hz and each subsequent waveform reduced by 15%. The polarity of successive AP waveforms alternated between positive and negative. This configuration was replicated across the eight AP channels. Similarly, in each PP channel, a chain of three equally spaced PP waveforms was placed, with the first waveform having an amplitude of 46 Hz and the second reduced by 30%.

Next, FID signals were generated using Equations (1) and (2) for a uniform sensing volume $\Delta V$, with representative MRI acquisitions parameters matching those in this study: sampling interval $\Delta t$ = 0.0977 ms, number of samples *NS* = 1024, echo time *TE*=15 ms, FID decay time constant $T_2$* = 50 ms, and *SNR* = 250 at *TE*. Gaussian random noise was added independently to the complex FID signals (real and imaginary components) across the channels *NCha* = 16 and between the real and imaginary parts, with a single noise trial *NScan* = 1. Noise was generated in MATLAB using the function *randn*(*NS*, *NCha*, *NScan*) to create normalized random noise matrices for the real and imaginary components separately.

The neuronal magnetic field component $B_{n,z}$ was then estimated from the noisy FID signals using Equations (3) and (4), as shown in Figure 2b. Measurement accuracy was quantified by comparing the amplitudes of the estimated firing peaks (Figure 2b) with their true values (Figure 2a).

<u>*Neuronal firing source localization in qsMRI measurements*</u>

Two approaches were used in this study to localize neuronal firing sources using coil element-specific MRI images. The first approach utilized standard anatomic imaging, while the second leverages images acquired during $B_0$ field shimming. Both approaches were implemented on a clinical 3T MRI scanner (Prisma Fit, Siemens MAGNETOM) equipped with a 20-channel Head/Neck array coil, comprising 16 head elements (H11–H44) and four neck elements (N11–N22).

In the anatomical imaging approach, two pulse sequences were employed. The first was a standard localizer (scout) sequence with additional prescriptions of three slices in each of three orthogonal planes, with uncombined (coil element-specific) images saved. These element-specific images defined qsMRI sensing volumes in the brain, such as those used in the phantom study (Figure 4a). The phantom consisted of a standard 2000-mL plastic bottle filled with saline water (0.5% ww in NaCl) and was equipped by MRI vendor (Siemens).

The second pulse sequence was a custom-developed whole-brain imaging sequence based on ultra-short echo time (UTE) with acquisition-weighed stack of spirals (AWSOS).[28] Although primarily used for other imaging purposes in the same study, the three-dimensional (3D) AWSOS sequence was leveraged here for qsMRI localization. Acquisition parameters were as follows: FOV = 220 mm; 60 slices with 3mm thickness without a gap between slices; 3D matrix size = 256×256×60; TE/TR = 0.5/80 ms; flip angle = 30º; 24 spiral interleaves per slice encoding; fat and spatial saturations; and total acquisition time (TA) = 1min56s. Offline image reconstruction was performed to generate coil element-specific images. Care was taken to correctly match image channels with FID signal channels using coil element labels (H11–H44, N11–N22). The resulting coil element-specific whole-brain images are shown in Figure 5h.

In the $B_0$ field-shimming approach, the shimming procedure was leveraged to generate low-spatial-resolution whole-brain images for localization of qsMRI measurements. The *k*-space raw data produced during the final iteration of $B_0$ shimming – identified by the MRI system as *AdjFieldMap.dat* and acquired immediately prior to qsMRI FID acquisition – were downloaded for offline image reconstruction. A custom-developed MATLAB program was used to reconstruct coil element-specific images from these raw data, yielding a 3D image of matrix size 80×64×75, as shown in Figures 7 and 8.

<u>*Neuronal firing detection by qsMRI*</u>

Two approaches were employed in this study to acquire proton ($^1$H) FID signals for qsMRI neuronal firing detection. The first approach used the product sequence *AdjFreq* during resonance-frequency adjustment to obtain a single FID signal. Acquisition parameters were readout time $T_s$ = 100 ms with dwell time (sampling interval) $\Delta t$ = 0.0997 ms, echo time $TE$ = 15ms, and flip angle =10º. This approach was used to acquire FID signals shown in Figures 4 and 5. The second approach used the product sequence *fid* to continuously acquire multiple, repeated FID signals. Acquisition parameters were $T_s$ = 819.2 ms with $\Delta t$ = 0.2 ms under an oversampling factor of 2, *TE*/*TR* = 0.2/1500 ms, flip angle = 90º, number of averages = 64, and total acquisition time *TA* = 1min36s. This approach was used to acquire FID signals under task and resting conditions, as presented in Figures 6–8.

*Neuronal firing magnetic-field calculation in qsMRI*

Based on Equations (3) and (4) and the flowchart shown in Figure 1f, a custom-developed MATLAB program (R2021a, MathWorks, Natick, MA) was used to calculate neuronal firing magnetic-field component $B_{n,z}$ from required FID signals. First, raw FID signal $s(t)$ was down-sampled to remove default oversampling applied by the MRI system during acquisitions. Down-sampling was performed by averaging two neighboring raw data points, rather than discarding every other data point, to improve signal-to-noise ratio. Next, phase difference $\Delta\varphi$ between adjacent time points was calculated by computing both a forward phase difference, $\Delta\varphi_R$ = *phase* $[s(t+\Delta t)s^*(t)]$, and a backward phase difference, $\Delta\varphi_L$ = *phase* $[s(t)s^*(t-\Delta t)]$. These were then averaged to yield final phase difference, $\Delta\varphi = (\Delta\varphi_R + \Delta\varphi_L)/2$. The phase difference was subsequently converted into neuronal magnetic-field component according to $B_{n,z} = \Delta\varphi/\gamma\Delta t$, where $\gamma$ is the gyromagnetic ratio. Three-point temporal smoothing after phase-difference calculation was applied in the analyses shown in Figures 3–8. Oversampling was retained in Figures 3–7 to enable large-range detection but was removed in Figure 8 to provide a cleaner visualization of firing peaks.

Phase unwrapping is generally unnecessary for FID signal $s(t)$ in Equation (3) because phase difference $\Delta\varphi$ between neighboring samples ($\pm\Delta t$) is computed from the phase of the conjugate product of adjacent complex samples. When sampling interval $\Delta t$ is sufficiently small – specifically, $\Delta t < \pi/(\gamma B_{n,z,max})$, where $B_{n,z,max}$ defines dynamic range $\pm B_{n,z,max}$ of qsMRI for neuronal magnetic field $B_{n,z}$ – the phase difference remains within [-$\pi, \pi$) and phase wrapping does not occur.

Quantum Sensing MRI for Detecting Neuronal Electrical Activity in Humans

# Supplementary Information

Yongxian Qian,[1,2,3,*] Ying-Chia Lin,[1,2] Sara Hejazi,[1,2] Kamri Clarke,[1,2] Kennedy Watson,[1,2] Xingye Chen,[1,4] Nahbila-Malikha Kumbella,[1] Justin Quimbo,[1] Abena Dinizulu,[5] Simon Henin,[6] Yulin Ge,[1,2] Arjun Masurkar,[5] Anli Liu,[6] Yvonne W. Lui,[1,2] Fernando E. Boada[1,2,†]

[1] Bernard and Irene Schwartz Center for Biomedical Imaging, Department of Radiology, New York University Grossman School of Medicine, New York, New York 10016, USA.

[2] Center for Advanced Imaging Innovation and Research (CAI2R), Department of Radiology, New York University Grossman School of Medicine, New York, New York 10016, USA.

[3] Neuroscience Institute, Department of Neuroscience and Physiology, New York University Grossman School of Medicine, New York, New York 10016, USA.

[4] Vilcek Institute of Graduate Biomedical Sciences, New York University Grossman School of Medicine, New York, New York 10016, USA.

[5]Alzheimer's Disease Research Center, Department of Neurology, NYU Langone Health, New York, New York 10016, USA

[6] Comprehensive Epilepsy Center, Department of Neurology, NYU Langone Health, New York, New York 10016, USA.

† Present address: Department of Radiology, Stanford University, Stanford, California 94305, USA

* **Corresponding author**. Email: Yongxian.Qian@nyulangone.org

## A. Computation scripts in MATLAB

### *A.1. Oversampling removal of FID signals via neighboring-sample averaging*

Consider an FID signal $s(t) = a(t)e^{j\varphi(t)}$ acquired at a channel with sampling interval (dwell time) $\Delta t$ and an oversampling factor of 2. Instead of discarding every other sample to remove oversampling, the signal-to-noise ratio (SNR) can be improved by averaging neighboring samples. Specifically, the oversampling removal is implemented as vector operation below,

$$s_2(1{:}NS/2) = (\ s(1{:}2{:}NS-1) + s(2{:}2{:}NS)\ )/2 \tag{S1}$$

where *NS* denotes total number of acquired data samples, and $s_2$ is the resulting downsampled FID signal. Alternatively, oversampling removal can be performed by first applying a lowpass filter to original FID signal and subsequently discarding every other sample.

### *A.2. Computation of phase difference between neighboring samples*

Forward and backward phase differences are computed and averaged as follows:

```
dPhaseBW = angle( s(1:NS).*conj(s([1,1:NS-1])) );     % backward
dPhaseFW = angle( conj(s(1:NS)).*(s([2:NS, NS])) );   % forward
dPhase = (dPhaseFW + dPhaseBW)/2;                     % average
```

### *A.3. Computation of neuronal magnetic field*

In presentation of frequency (Hz): `nfreq = dPhase/(2*pi*dt);`
In presentation of magnetic field (Tesla): `Bnz = nfreq/gammaBar;`
With `dt` $= \Delta t$, `pi` $= \pi$, and `gammaBar` $= \bar{\gamma}$.

### *A.4. Counting number of neuronal firing peaks*

Threshold neuronal magnetic field Bnz in unit of mirco Tesla (uT)

```
BnzMag = abs(Bnz);                   % take magnitude
BnzMag(BnzMag < epochAmpl) = 0;      % threshold upon predefined epochAmpl=1uT
BnzMag(NT+1:end,:,:)       = 0;      % cut-off tail after NTth sample
```

Detect firing peaks through the three-point (left, mid, right) comparison

```
NFPeaks = 0*BnzMag;                  % initiate labeling matrix for NF peaks
for jScan=1:NScans
    for jCha=1:NCha
        for jSample=2:NT-1
            a1 = BnzMag(jSample-1, jCha,jScan);
            a2 = BnzMag(jSample  , jCha,jScan);
            a3 = BnzMag(jSample+1, jCha,jScan);
            if 2*a2 > a1+a3
                NFPeaks(jSample,jCha,jScan) = 1;      % label as a peak
            end
        end % end of jSample loop
    end % end of jCha loop
end % end of jScan loop
```

Calculate firing rate, mean, and SD

```
NFRate = squeeze(sum(NFPeaks,1)./(0.001*epochMs);% Hz,peaks/s,[NCha,NScans]
NFRateMean  = squeeze(mean(NFRate,2));           % Hz,[NCha],along scans/TRs
NFRateSD    = squeeze(std(NFRate,0,2));          % Hz,[NCha],along scans/TRs
```

### *A.5. Simulation of action potentials (APs) and postsynaptic potentials (PPs or PSs)*

Key functional segments of the MATLAB computational code are provided here to facilitate independent reproducibility and to enable customized simulations with user-defined parameters. Intermediate outputs are also listed in Table A4 for confirmation.

Set simulation parameters:

```
NS=1024; NCha=16; NScan=1; snr=250; % s/SD
```

```
TE=15; T2=50;       % millisecond, ms
dts=0.0977e-3;      % second, s
gammaBar = 42.58; % MHz/T for proton (1H) nucleus.
```

Get the predesigned AP and PP (PS) waveforms:

```
load ('fAP-PS.mat','fAP','fPS'); % fAP and fPS in micro Tesla, uT
```

Create AP/PP channels:

```
fAP2=[fAP;fAP]; fAP2(1:2:NS-1)=fAP/5; fAP2(2:2:NS)=fAP/5; % adjust amplitude
fPS2=[fPS;fPS]; fPS2(1:2:NS-1)=fPS/1; fPS2(2:2:NS)=fPS/1;
```

Create time axis:

```
tms=TEms+((1:NS)'-1)*dts*1000;       % ms
```

Calculate phase via Equation (1):

```
phaAP =cumsum(fAP2);     % cumulative sum
phaPS =cumsum(fPS2);
```

Calculate function exp():

```
expAP = exp(2i*pi*dts*phaAP*gammaBar);     % i = sqrt(-1)
expPS = exp(2i*pi*dts*phaPS*gammaBar);
```

Create T2 decay: `decay = exp(-tms/T2);`

Generate normalized complex random noise, independent from channel to channel

```
noiseReal = randn(NS,NCha,NScan);
noiseImag = randn(NS,NCha,NScan);
noiseData = noiseReal+ 1i*noiseImag;       % complex-valued noise
```

Simulate FID signals using practical amplitudes from a 25-year-old healthy female subject:

```
load ('AmplRawData.mat','AmplRawData');    % practical amplitude values
fidSimu = 0*noiseData;                     % declare fidSimu as complex data
for iCha=1:2:NCha                          % interleaved PP/AP channels
    fidSimu(:,iCha  ) = AmplRawData(iCha).*decay.*expPS;
    fidSimu(:,iCha+1) = AmplRawData(iCha).*decay.*expAP;
end
```

Add random noise to simulated FID signals upon SNR:

```
fidSimuNoisy = fidSimu;                    % declare noisy fid signal
for iCha=1:2:NCha
      fidSimuNoisy(:,iCha  ) = fidSimu(:,iCha) + …
                          (AmplRawData(iCha)/snr)*noiseData(:,iCha,1);
      fidSimuNoisy(:,iCha+1) = fidSimu(:,iCha+1) + …
                          (AmplRawData(iCha)/snr)*noiseData(:,iCha+1,1);
end
```

**Table A4. Intermediate outcomes of the simulations above.**

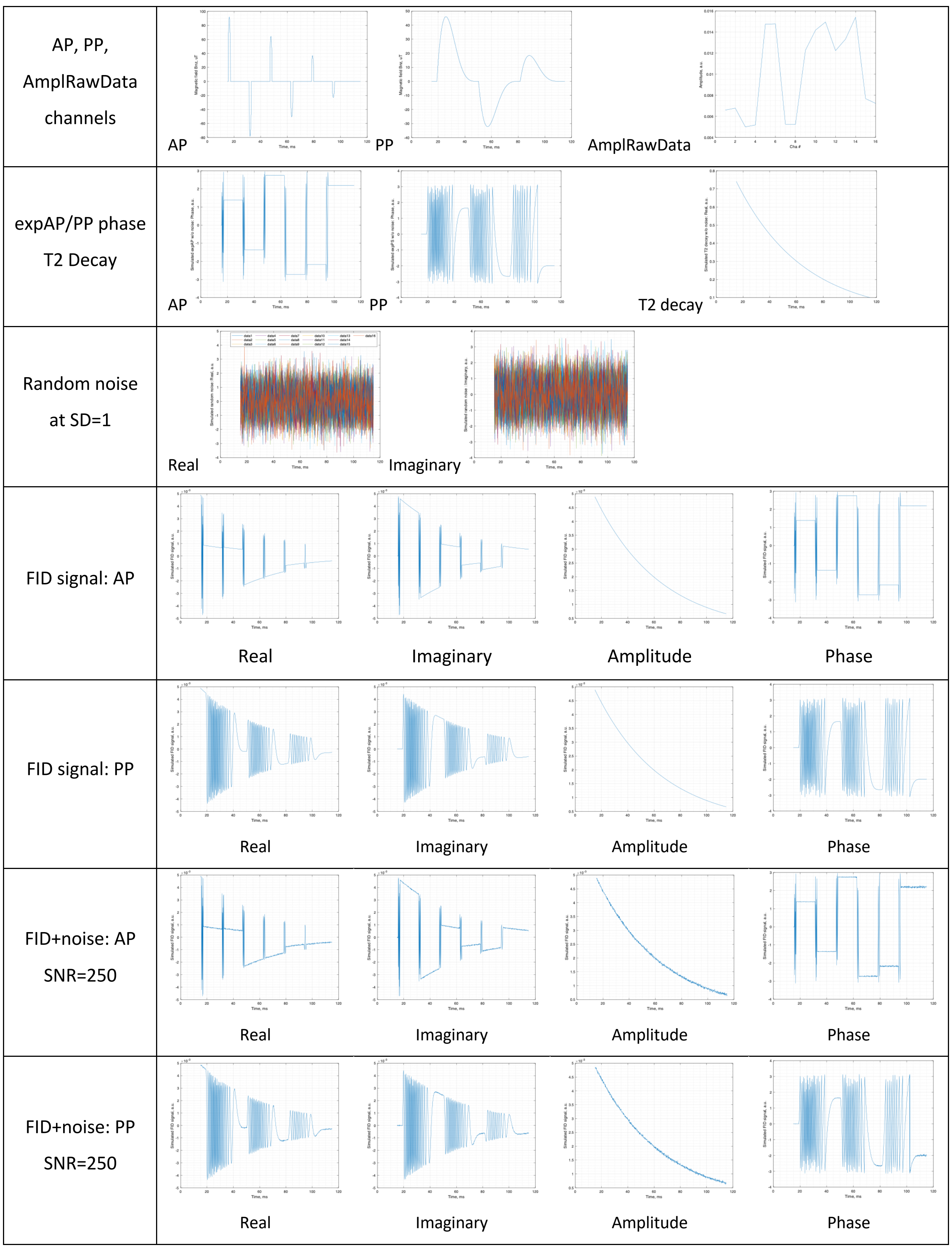

## B. Accuracy and precision of qsMRI measurements based on numerical simulations in Figure 2

The results in Tables B1 and B2 are based on the simulations at SNR=250 – a practical level from our human brain studies.

**Table B1. Accuracy of qsMRI measurements at SNR = 250**

| AP peak, µT | | | PP peak, µT | | |
|---|---|---|---|---|---|
| True | Measured | Difference, % | True | Measured | Difference, % |
| -78.2 | -77.4318 | -0.9823529 | -32.1837 | -32.189075 | 0.016701 |
| -49.0432 | -50.02475 | 2.00139877 | 18.3907 | 18.3326425 | -0.3156895 |
| -23 | -22.647325 | -1.5333696 | 45.9704 | 45.986225 | 0.03442433 |
| 35.6678 | 36.4709175 | 2.25165976 | | | |
| 64.4 | 63.612375 | -1.2230202 | | | |
| 92 | 91.13975 | -0.9350543 | | | |
| **Average** | | **-0.1714412** | **Average** | | **-0.088188** |

**Table B2. Precision of qsMRI measurements at SNR = 250**

| Measured value of AP peaks, 2x µT | | | | | | | | MEAN | SD | $\delta$, % |
|---|---|---|---|---|---|---|---|---|---|---|
| -38.7355 | -38.7737 | -38.761 | -38.697 | -38.672 | -38.663 | -38.743 | -38.682 | -38.7159 | 0.04263711 | -0.11013 |
| -25.1336 | -24.9314 | -25.113 | -25.095 | -24.915 | -24.799 | -24.958 | -25.154 | -25.012375 | 0.1288575 | -0.51518 |
| -11.4414 | -11.2669 | -11.347 | -11.405 | -11.222 | -11.341 | -11.304 | -11.262 | -11.323663 | 0.07470825 | -0.65975 |
| 18.10647 | 18.3627 | 18.1915 | 18.328 | 18.195 | 18.389 | 18.158 | 18.153 | 18.2354588 | 0.10779333 | 0.591119 |
| 31.7176 | 31.7646 | 31.9143 | 31.784 | 31.871 | 31.869 | 31.775 | 31.754 | 31.8061875 | 0.06928217 | 0.217826 |
| 45.6212 | 45.5431 | 45.6141 | 45.5016 | 45.593 | 45.57 | 45.499 | 45.617 | 45.569875 | 0.05035204 | 0.110494 |
| | | | | | | | | | **Average** | **-0.06094** |
| Measured value of PP peaks, 2x µT | | | | | | | | | | |
| -16.1864 | -16.0344 | -16.0995 | -16.09 | -16.209 | -16.034 | -16.087 | -16.016 | -16.094538 | 0.07080341 | -0.43992 |
| 9.25837 | 9.1628 | 9.0564 | 9.364 | 9.37 | 9.155 | 8.853 | 9.111 | 9.16632125 | 0.17007931 | 1.855481 |
| 23.0581 | 22.9474 | 22.9996 | 23.0028 | 22.933 | 22.981 | 23.004 | 23.019 | 22.9931125 | 0.03963741 | 0.172388 |
| | | | | | | | | | **Average** | **0.529316** |

**Table C1. Matching coil elements between source-localizing images and FID signal channels in the study subject with a history of epilepsy shown in Figure 8.**

| AdjFieldMap imaging | | | | | | FID signals | | | | | |
|---|---|---|---|---|---|---|---|---|---|---|---|
| asList [] in scanHeader | Element | ADC Cha Connected | ADC Cha Sorted | **Element After Sort** | Cha # in RawData | asList [] in scanHeader | Element | ADC Cha Connected | ADC Cha Sorted | **Element after Sort** | Cha # in RawData |
| 0 | H31 | 30 | 17 | **N12** | 1 | 0 | S11 | 50 | 17 | **H14** | 1 |
| 1 | H32 | 29 | 18 | **N11** | 2 | 1 | S12 | 49 | 18 | **H13** | 2 |
| 2 | H33 | 50 | 21 | **H14** | 3 | 2 | S13 | 42 | 21 | **H23** | 3 |
| 3 | H34 | 49 | 22 | **H13** | 4 | 3 | S14 | 41 | 22 | **H24** | 4 |
| 4 | H41 | 38 | 25 | **H23** | 5 | 4 | H11 | 30 | 25 | **H32** | 5 |
| 5 | H42 | 37 | 26 | **H24** | 6 | 5 | H12 | 29 | 26 | **H31** | 6 |
| 6 | H43 | 42 | 29 | **H32** | 7 | 6 | H13 | 18 | 29 | **H12** | 7 |
| 7 | H44 | 41 | 30 | **H31** | 8 | 7 | H14 | 17 | 30 | **H11** | 8 |
| 8 | H21 | 61 | 33 | **H12** | 9 | 8 | H21 | 53 | 33 | **H42** | 9 |
| 9 | H22 | 62 | 34 | **H11** | 10 | 9 | H22 | 54 | 34 | **H41** | 10 |
| 10 | H23 | 25 | 37 | **H42** | 11 | 10 | H23 | 21 | 37 | **H44** | 11 |
| 11 | H24 | 13 | 38 | **H41** | 12 | 11 | H24 | 22 | 38 | **H43** | 12 |
| 12 | H11 | 34 | 41 | **H44** | 13 | 12 | H41 | 34 | 41 | **S14** | 13 |
| 13 | H12 | 33 | 42 | **H43** | 14 | 13 | H42 | 33 | 42 | **S13** | 14 |
| 14 | H13 | 22 | 45 | **S14** | 15 | 14 | H43 | 38 | 45 | **H34** | 15 |
| 15 | H14 | 21 | 46 | **S13** | 16 | 15 | H44 | 37 | 46 | **H33** | 16 |
| 16 | S11 | 54 | 49 | **H34** | 17 | 16 | H31 | 26 | 49 | **S12** | 17 |
| 17 | S12 | 53 | 50 | **H33** | 18 | 17 | H32 | 25 | 50 | **S11** | 18 |
| 18 | S13 | 46 | 53 | **S12** | 19 | 18 | H33 | 46 | 53 | **H21** | 19 |
| 19 | S14 | 45 | 54 | **S11** | 20 | 19 | H34 | 45 | 54 | **H22** | 20 |
| 20 | **N21** | 58 | 57 | **N22** | 21 | | | | | | |
| 21 | **N22** | 57 | 58 | **N21** | 22 | H11-H44: | Head elements | | | | |
| 22 | **N11** | 18 | 61 | **H21** | 23 | **N11-N22:** | **Neck elements** | | | | |
| 23 | **N12** | 17 | 62 | **H22** | 24 | **S11-S14:** | **Spinal elements** | | | | |

## C. Matching coil elements to ADC channels between FID signals and source-localizing images

The MRI system used in this study (Prisma Fit, version VE11C, Siemens Healthneers) employs dynamic assignment of receive coil elements to analog-to-digital (ADC) channels. Consequently, the mapping between a specific individual coil element and its corresponding ADC channel may vary across imaging protocols (pulse sequences). Although this dynamic assignment is recorded in the scan header, it is not stored in the channel header of the raw *k*-space data. This discrepancy introduces ambiguity when matching coil elements between FID acquisitions and source-localizing images, as the ADC channel indices in the raw data do not directly reflect a fixed coil-element identity. To ensure proper alignment between image channels and FID signal channels, we implemented a manual matching procedure as described below. This manual match will be automated in future when inline display is implemented. Table C1 presents an example how to match the image-FID channels via coil elements.

## D. Measurement of neuronal firing peaks (singlets) in Figure 3.

Three representative neuronal firing peaks (singlets) were measured from a subject's qsMRI recording, as shown in Figuire 3 (Channel 7), to assess their temporal proximity to a single-neuron firing AP event (~2 ms). Quantitative measurements are provided in Table D1.

**Table D1. Measurement of singlet peak duration**

| Peak | t1 (start), ms | t2 (end), ms | Duration, ms |
| --- | --- | --- | --- |
| Singlet 1 | 63.4592 | 64.436 | 0.9768 |
| Singlet 2 | 73.8154 | 75.3786 | 1.5632 |
| Singlet 3 | 78.1142 | 79.6774 | 1.5632 |
| **Average** | | | **1.3677** |

## E. Experimental protocols for the open validation

*<u>E1. A suitable ground-truth model</u>*

It can generate action potentials and/or postsynaptic potentials with controlled waveform, duration, and firing frequencies, and transmit these signals along myelinated axons with well-defined nodes of Ranvier. At each node, paired dipolar electric currents are produced within the axon, flowing in opposite directions – an essential feature of neuronal signal propagation. These paired currents generate neuronal magnetic field both inside and outside the axon: a dominant dipolar field within the axon and a much weaker quadrupolar magnetic field outside the axon. qsMRI is sensitive specifically to the intra-axon dipolar magnetic field, particularly its component Bn,z aligned with the main static field B0. Importantly, the microscopic ionic environment within the axon must be significantly inhomogeneous. Sufficient uniform would lead to cancellation of the magnetic-field effects on proton nuclear spins, rendering the neuronal magnet field undetectable by qsMRI.

*E2. FID signal acquisition*

MRI experiments should be performed on a scanner operating at 3T or higher to ensure sufficient SNR in the FID signals. The receiver coil may be a multi-channel array to enable localization of firing sources, or a single-channel volume coil combined with selective RF excitation for targeting a specific region of the model. Manual B0 field shimming should be performed to achieve a water linewidth below 30 Hz. Data acquisition should use either *fid* or single-voxel spectroscopy (*SVS*) pulse sequence without water suppression. FID sampling interval (including oversampling) must be ≤ 0.1 ms for an 819.2ms-long readout, and at least 64 FID acquisitions should be collected.

*E3. Firing source localization*

when an *fid* sequence is used in combination with a multi-channel array coil, coil-specific MRI images are required for localization of neuronal firing sources. These images may be obtained through separate anatomical imaging acquisitions (e.g., MPRAGE, FLAIR, or $T_2$-weighted imaging) or by reconstructing images from the $B_0$-shimming data acquired immediately prior to the FID measurements. Image reconstruction may be performed either inline or offline, as demonstrated in Figure 6c.

*E4. Data processing*

Follow the flowchart in Figure 1f. Specifically, (i) Preprocessing. FID signals should first be visually inspected to assess the impact of random noise and to determine an appropriate analysis time window (e.g., 300 or 500 ms) for firing rate calculation. Frequency spectra should also be examined for off-resonance water peak shifts and fat peak interference. If necessary, water peak should be corrected for off-resonance shift and fat peak removed. When random noise is substantial, FID signals may be downsampled from 0.1 ms to 0.2 or 0.3 ms by averaging two or three neighboring data points. (ii) Neuronal magnetic field estimation. The neuronal magnetic field component $B_{n,z}$ should then be calculated from the (optionally down-sampled) FID signals using Equations (3)–(4). (iii) Firing-rate calculation. neuronal firing rates are computed by counting firing peaks within the predetermined time window and normalizing by the window duration.

// *The end of Supplementary Information*